\documentclass[journal]{IEEEtran}
\usepackage{academicons}
\usepackage{amsmath}
\usepackage{booktabs}
\usepackage{hyperref}
\usepackage{multirow}
\usepackage{nicefrac}
\usepackage{xcolor}
\newcommand{\orcid}[1]{\href{https://orcid.org/#1}{\includegraphics[width=8pt]{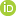}}}

\usepackage{amsfonts}
\usepackage{cite}

\ifCLASSINFOpdf
  \usepackage[pdftex]{graphicx}
\else
\fi
\begin{document}
%
\title{HiddenSinger: High-Quality Singing Voice Synthesis via Neural Audio Codec and Latent Diffusion Models}
%
%
\author{Ji-Sang~Hwang\orcid{0000-0002-0361-9939},
        Sang-Hoon~Lee\orcid{0000-0002-8925-4474},
        and Seong-Whan~Lee\orcid{0000-0002-6249-4996} ,~\IEEEmembership{Fellow,~IEEE}
\thanks{This work was supported by Institute of Information \& communications Technology Planning \& Evaluation (IITP) grant funded by the Korea government (MSIT) (No. 2019-0-00079, Artificial Intelligence Graduate School Program (Korea University) and No. 2021-0-02068, Artificial Intelligence Innovation Hub) and Netmarble AI Center. \textit{(Corresponding author: Seong-Whan Lee.)} Ji-Sang Hwang, Sang-Hoon Lee, and Seong-Whan Lee are with the Department of Artificial Intelligence, Korea University, Seoul 02841, South Korea (e-mail: js\_hwang@korea.ac.kr;sh\_lee@korea.ac.kr;sw.lee@korea.ac.kr).}
}

\maketitle

\begin{abstract}
Recently, denoising diffusion models have demonstrated remarkable performance among generative models in various domains. However, in the speech domain, the application of diffusion models for synthesizing time-varying audio faces limitations in terms of complexity and controllability, as speech synthesis requires very high-dimensional samples with long-term acoustic features. To alleviate the challenges posed by model complexity in singing voice synthesis, we propose HiddenSinger, a high-quality singing voice synthesis system using a neural audio codec and latent diffusion models. To ensure high-fidelity audio, we introduce an audio autoencoder that can encode audio into an audio codec as a compressed representation and reconstruct the high-fidelity audio from the low-dimensional compressed latent vector. Subsequently, we use the latent diffusion models to sample a latent representation from a musical score. In addition, our proposed model is extended to an unsupervised singing voice learning framework, HiddenSinger-U, to train the model using an unlabeled singing voice dataset. Experimental results demonstrate that our model outperforms previous models in terms of audio quality. Furthermore, the HiddenSinger-U can synthesize high-quality singing voices of speakers trained solely on unlabeled data.
\end{abstract}


\ifCLASSOPTIONpeerreview
\begin{center} \bfseries EDICS Category: 3-BBND \end{center}
\fi
%
\IEEEpeerreviewmaketitle

\section{Introduction}
\label{sec:introduction}
\begin{figure}[ht!]
    \includegraphics[width=\linewidth]{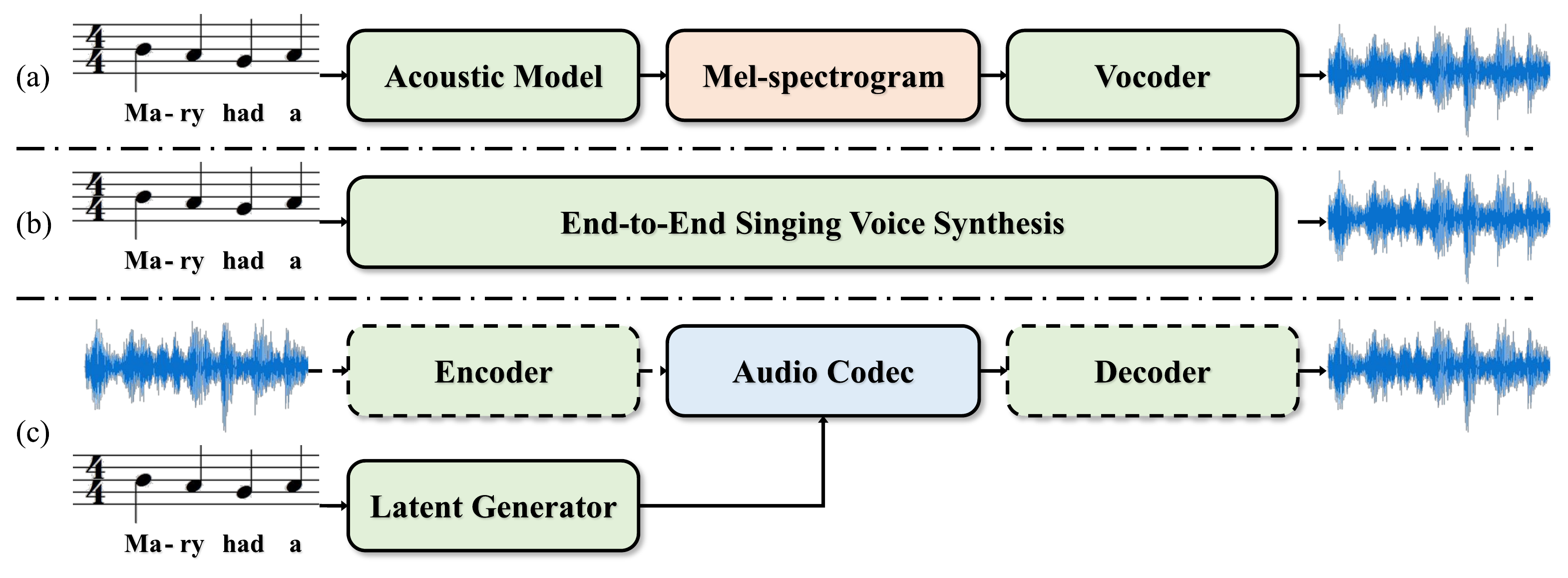}
    \caption{Comparison of SVS system architectures: (a) two-stage pipeline SVS system with pre-defined intermediate representation; (b) end-to-end SVS system; (c) proposed SVS system that uses an audio codec from the pre-trained audio autoencoder. The dashed outlines indicate that the parameters of the models are not updated during the generation of the audio codec from a musical score and the synthesis of audio from the audio codec.}\vspace{-0.2cm}
    \label{fig:svs_overview}
    \vspace{-0.2cm}
\end{figure}

\IEEEPARstart{S}{inging} voice synthesis (SVS) systems aim to generate high-quality expressive singing voices from musical scores. Recent advancements in generative models \cite{kingma2013auto,rezende2015variational,goodfellow2020generative} have led to rapid development in deep-learning-based SVS systems, resulting in high performance. Most SVS systems first synthesize an intermediate acoustic representation, such as Mel-spectrogram, from a musical score using an acoustic model \cite{degottex2016multi,chen2020hifisinger,hono2021sinsy,lee2021n,choi2022melody,liu2022diffsinger}. Subsequently, separately trained vocoders \cite{chen2021singgan,wu2022ddsp,lee2022duration} convert the generated representation into audio, as shown in Fig \ref{fig:svs_overview}(a).

However, conventional two-stage SVS systems face certain limitations. These systems depend on pre-defined intermediate representation, making it difficult to apply latent learning to improve audio generation. Moreover, a training-inference mismatch problem occurs because the predicted intermediate representation differs from the ground-truth intermediate representation. To resolve these issues, an end-to-end SVS system, VISinger \cite{zhang2022visinger}, directly synthesizes audio by employing variational inference.

Although existing systems can improve the audio quality, several challenges remain: 1) SVS systems require high-dimensional audio or a linear-spectrogram to synthesize high-fidelity audio, resulting in high computational costs in high-dimensional space. 2) The training-inference mismatch problem persists in end-to-end systems. A gap between the posterior distribution from the audio and the prior distribution from the musical score exists, which results in inaccurate pitch and mispronunciations in the generated singing voice. Moreover, systems based on Normalizing Flows \cite{rezende2015variational} are trained in a backward direction but perform inference in a forward direction \cite{tan2022naturalspeech}. 3) SVS systems require audio-musical score corpora for training, wherein it is time-consuming to obtain high-quality paired datasets.

To address the aforementioned problems, we propose HiddenSinger, an advanced high-quality SVS system utilizing a neural audio codec and latent diffusion models. Our approach involves multiple components to enhance the synthesis process: First, we introduce an audio autoencoder that can efficiently encode audio into a compressed latent representation, resulting in a lower-dimensional representational space. We also adopt residual vector quantization in the audio autoencoder to regularize the arbitrarily high-variance latent space. Subsequently, we employ the powerful generative ability of latent diffusion models to generate a latent representation conditioned on a musical score, which is converted into audio through the audio autoencoder. Moreover, we propose an unsupervised singing voice learning framework that leverages unpaired singing voice data containing only audio. The experimental results demonstrate that HiddenSinger outperforms previous SVS models in terms of audio quality. Furthermore, our model can synthesize high-quality singing voices, even for speakers who are represented in unpaired data, by using the proposed unsupervised singing voice learning framework (HiddenSinger-U).

Our study makes the following contributions:
\begin{itemize}
    \item We introduce HiddenSinger, which utilizes a neural audio codec and latent diffusion models to synthesize high-quality singing voices. The latent generator generates a latent representation conditioned on a musical score. Subsequently, the audio autoencoder synthesizes high-quality singing voice audio from the generated latent representation.
    \item We extend our proposed model to HiddenSinger-U, an unsupervised singing voice learning framework that performs training with both paired and unpaired datasets using acoustic features from audio. HiddenSinger-U can synthesize a high-quality singing voice of a speaker without a musical score during training.
    \item The proposed model is demonstrated to outperform previous SVS models. Audio samples are available at \url{https://jisang93.github.io/hiddensinger-demo/}
\end{itemize}

\section{Related Studies}

\subsection{Singing Voice Synthesis}
\label{subsec:svs}
Singing voice synthesis (SVS) systems are designed to generate a singing voice based on a musical score. Since singing voices comprise significant pitch variability and an extended duration of vowel, SVS systems require additional input data, such as note pitch, note duration, and lyrics. Conventional SVS systems follow a two-stage manner comprising an acoustic model \cite{chen2020hifisinger,hono2021sinsy,lee2021n,choi2022melody,liu2022diffsinger} and a vocoder \cite{chen2021singgan,wu2022ddsp} to synthesize a realistic singing voice. Although previous SVS systems improved the singing voice quality, the two-stage pipeline has inherent limitations that prevent it from surpassing the upper bound of the vocoder performance.

To address these limitations, researchers have proposed end-to-end SVS systems \cite{zhang2022visinger} that use a well-learned latent representation to enhance the quality of singing voices and simplify the training procedure. However, the end-to-end method still faces problems, including a training-inference mismatch problem. Specifically, the gap between the posterior and prior distributions leads to degraded audio reconstruction performance. In this study, we leverage a well-learned latent representation, which is converted into an audio codec, to improve the quality of the reconstructed audio.

\subsection{Neural Audio Synthesis}
To generate natural audio, neural vocoders \cite{oord2016wavenet,yamamoto2020parallel,ai2020neural} are generally used to convert signal processing components, such as the Mel-spectrogram, into raw waveform audio. For high-quality audio generation, a generative adversarial network (GAN)-based neural vocoder adopts a multi-scale discriminator \cite{kumar2019melgan} and a multi-period discriminator \cite{kong2020hifi} to capture the specific characteristics of the waveform audio. Although a diffusion-based neural vocoder \cite{huang2022fastdiff} has been presented, several limitations persist in the audio quality and inference speed in tasks concerning waveform audio generation.

In recent developments, neural audio codecs have emerged in conjunction with neural vocoders. These audio codecs efficiently compress the audio in an autoencoder architecture. For improved compression, approaches such as SoundStream \cite{zeghidour2021soundstream} introduce residual vector quantization, leading to enhanced coding efficiency. Encodec \cite{defossez2022high} also represents the audio as discrete units with the residual vector quantization and incorporates a multi-scale short-time Fourier transform (STFT)-based discriminator to reduce artifacts in the reconstructed audio. Drawing inspiration from these studies, we adopt neural audio codecs to achieve high-fidelity audio generation and computational efficiency.

\subsection{Diffusion Probabilistic Model}
\label{subsec:diffusion}
Diffusion probabilistic models (also known as diffusion models) \cite{ho2020denoising} are a class of generative models that have achieved remarkable results in various domains, such as image \cite{dhariwal2021diffusion,ramesh2022hierarchical,rombach2022high}, audio \cite{yang2023diffsound,huang2023make} and video \cite{ho2022video} generation. Particularly, in the audio domain, previous studies have mainly used diffusion models to generate acoustic features.

For the acoustic feature generation, Grad-TTS \cite{popov2021grad}, DiffSinger \cite{liu2022diffsinger}, and DDDM-VC \cite{choi2023dddm} utilize the diffusion-based decoder to generate a high-quality Mel-spectrogram. Each model uses a conditional-diffusion decoder to condition text distribution for a text-to-speech system \cite{popov2021grad}, the musical score for a SVS system, and speaker information for a voice conversion system. To improve the generation efficiency by compressing the Mel-spectrogram into discrete latent space, DiffSound \cite{yang2023diffsound} introduces a discrete diffusion-based token decoder in a non-autoregressive manner. Make-an-audio \cite{huang2023make} adopts the latent diffusion models to generate a continuous latent representation that converts into Mel-spectrogram. For waveform generation, Diffwave \cite{kong2020diffwave} and WaveGrad \cite{chen2020wavegrad} generate high-fidelity speech waveform from the Mel-spectrogram. In contrast to the above approaches, WaveGrad 2 \cite{chen2021wavegrad} and FastDiff \cite{huang2022fastdiff} adopt an end-to-end manner that generates the audio without any intermediate features (e.g., Mel-spectrogram). Inspired by the success of diffusion-based generation, we adopt latent diffusion models to generate a latent representation conditioned on a musical score.

\section{Preliminary}
Diffusion models comprise two processes: a forward process (diffusion process) and reverse process (denoising process). In the forward process, the data $X_0$ are gradually corrupted with a tiny Gaussian noise through a $T$-step Markov chain. The reverse process, which follows the reverse trajectory of the forward process, aims to generate the data $X_0$ from the Gaussian noise $X_T$.

\begin{figure*}[ht]
    \includegraphics[width=\linewidth]{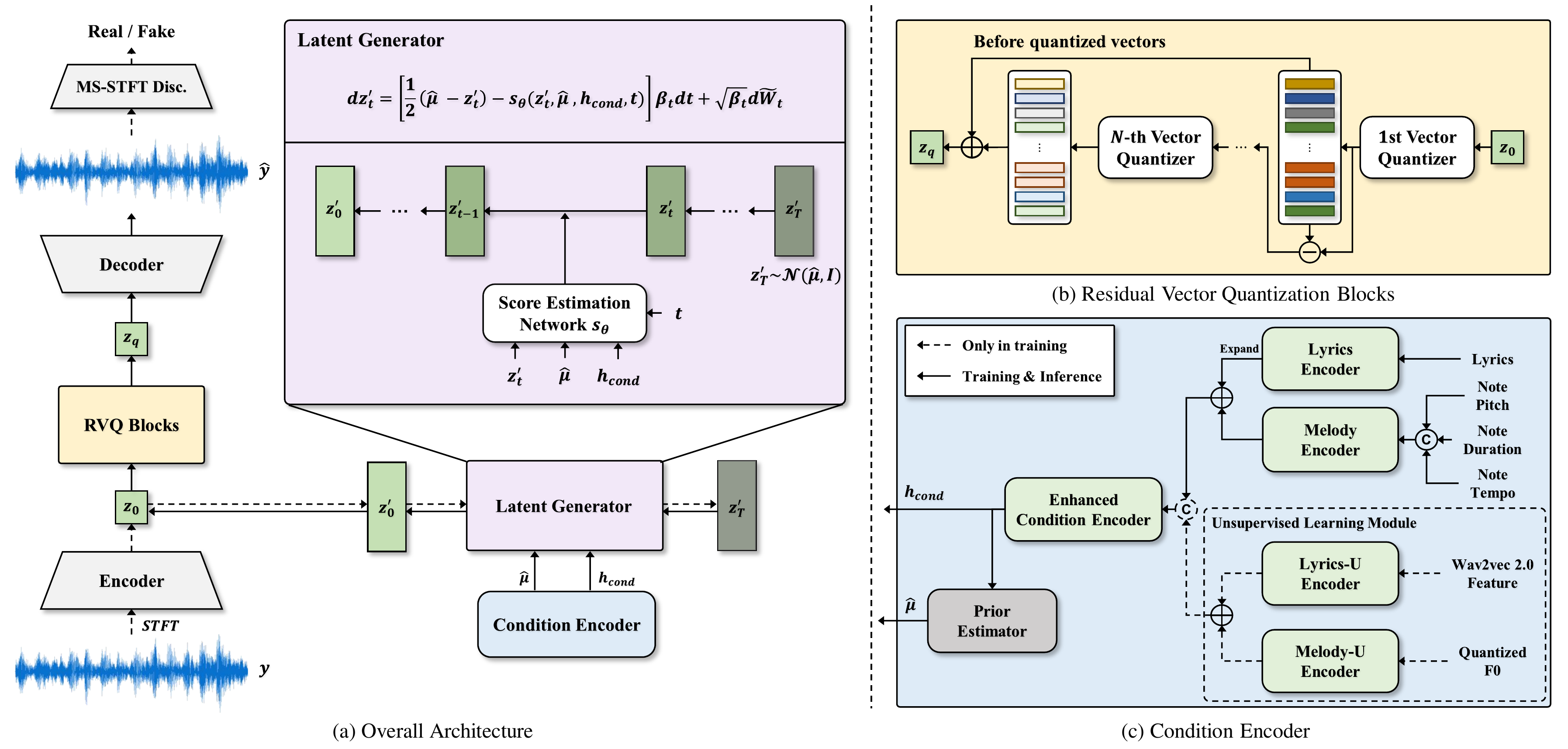}
    \caption{(a) Architecture of HiddenSinger. We train the audio autoencoder and latent generator separately. During the inference, the latent generator gradually denoises a noisy sample from the data-driven priors. Then, the audio autoencoder converts the sampled latent representation into audio. (b) The RVQ blocks discretize the continuous latent representation into an audio codec. (c) To guide the latent generator, the condition encoder extracts a condition representation $h_{cond}$ and estimates $\hat{\mu}$ from a musical score. The dashed arrows indicate that the operations are only used during training.}
    \label{fig:arch_overall}
    \vspace{-0.2cm}
\end{figure*}

In \cite{song2020score,vahdat2021score}, a stochastic differential equation (SDE) was used to approximate the trajectory between $X_0$ and $X_T$. In the speech domain, Grad-TTS \cite{popov2021grad} and Guided-TTS \cite{kim2022guided} applied an SDE to the text-to-speech task. Following \cite{song2020score}, forward process that perturbs the data $X_{0}$ into the noise $X_{T}$ is defined with the pre-defined noise schedule $\beta_{t}=\beta_{0}+(\beta_{T}-\beta_{0})t$:
\begin{align}
    dX_{t} = -\frac{1}{2}X_{t}\beta_{t}dt + \sqrt{\beta_{t}}dW_{t},
\end{align}
where $W_{t}$ represents the standard Brownian motion and $t$ denotes a continuous timestep $t\in[0, T]$.

The reverse process is defined as a reverse-time SDE that formulates the trajectory from Gaussian noise $X_T$ to the data $X_0$ as follows:
\begin{align}
\label{eq:2}
    dX_{t} = \left(-\frac{1}{2}X_{t} - \nabla_{X_{t}}\log{p_{t}(X_t)}\right)\beta_{t}dt + \sqrt{\beta_{t}}d\tilde{W}_{t},
\end{align}
where $\tilde{W}_{t}$ is the reverse Brownian motion and $\nabla_{X_{t}}\log{p_t(X_t)}$ represents a score of the probability density function of data $p_t(X_t)$.

A neural network $s_{\theta}$ learns to estimate the score, which is parameterized by $\theta$, to model the data distribution $p_t(X_t)$. By solving Eq. \ref{eq:2}, $X_{0} \sim p_{0}(X)$ can be obtained by starting from the noisy data $X_{T}$ and iteratively removing the noise using the score estimation networks $s_{\theta}$.

\section{HiddenSinger}
\label{sec:method}
In this paper, we propose a SVS system using neural audio codecs and latent diffusion for high-quality singing voice audio. We introduce an audio autoencoder using residual vector quantization to achieve high-fidelity audio generation and computational efficiency. Additionally, we adopt latent diffusion models in a latent generator to generate a latent representation conditioned on a musical score, which is converted into audio by the audio autoencoder. Furthermore, we extend HiddenSinger to HiddenSinger-U, which can train the model without musical scores. In the following subsection, we describe the details of HiddenSinger and an unsupervised singing voice learning framework (HiddenSinger-U).

\subsection{Audio Autoencoder}
\label{subsec:audio_autoencoder}
For efficient coding and high-quality audio generation, we introduce the audio autoencoder to compress the audio into an audio codec, which provides a low-dimensional representation. The audio autoencoder comprises three modules: an encoder, residual vector quantization (RVQ) blocks, and a decoder, as illustrated in Fig. \ref{fig:arch_overall} (a).

\subsubsection{Encoder}
The encoder takes a high-dimensional linear-spectrogram as the input and extracts a low-dimensional continuous latent representation $z_{0}$ from the audio $y$. Inspired by \cite{rombach2022high}, the latent space is regularized through vector quantization (VQ) \cite{van2017neural} to avoid an arbitrarily high-variance of the latent space. A previous study in which sampling was performed using latent diffusion models demonstrated that a model trained on the VQ-regularized latent space achieved better quality than the Kullback-Leibler (KL)-regularized latent space. In our preliminary experiments, we observed that the KL-regularized latent space achieved sub-optimal performance when the diffusion models restored the latent representation. However, conventional VQ is insufficient for high-fidelity audio reconstruction because a quantized vector should represent multiple features of a raw waveform. Therefore, we apply RVQ \cite{zeghidour2021soundstream} to the continuous latent representation $z_{0}$ for efficient audio compression.

\subsubsection{Residual Vector Quantization Blocks}
As indicated in Fig. \ref{fig:arch_overall} (b), the first vector quantizer discretizes the continuous latent representation $z_{0}$ into the closest entry in a codebook. Subsequently, the residual is computed. The next quantizer is used with the second codebook, with this process repeated as many times as the number of quantizers $C$. The number of quantizers is related to the trade-off between the computational cost and coding efficiency. We follow the training procedure described in \cite{zeghidour2021soundstream} to train the codebook for each quantizer. Furthermore, we apply the commitment loss \cite{van2017neural} to stabilize the codebook training. We found that the low-weighted commitment loss helps to converge the RVQ blocks during training:
\begin{align}
    \mathcal{L}_{emb}={\sum^{C}_{c=1}||z_{0, c}-q_{c}(z_{0, c})||_{2}^{2}},
\end{align}
where $z_{0, c}$ represents the residual vector of the $c$-th quantizer and $q_{c}(z_{0, c})$ denotes the closest entry in the $c$-th codebook.

\subsubsection{Decoder}
The decoder generates a raw waveform from the audio codec $\hat{y}=G{\left(z_{q}\right)}$. We calculate a reconstruction loss $\mathcal{L}_{recon}$ between the generated $\hat{x}_{mel}$ and ground-truth Mel-spectrograms $x_{mel}$ to improve the training efficiency of the decoder. The reconstruction loss is defined as
\begin{align}
    \mathcal{L}_{recon}={||x_{mel}-\hat{x}_{mel}||}_1.
\end{align}

Moreover, we adopt adversarial learning to improve the quality of the generated audio. We use a multi-scale STFT-based (MS-STFT) discriminator \cite{defossez2022high}, which expands a multi-resolution spectrogram discriminator \cite{jang2021univnet}. The MS-STFT discriminator operates on a multi-scale complex-valued STFT that contains both real and imaginary parts. Similar to the work of \cite{defossez2022high}, we observed that the MS-STFT discriminator trains the decoder efficiently and facilitates the synthesis of audio with better quality than the combination of a multi-period discriminator \cite{kong2020hifi} and multi-scale discriminator \cite{kumar2019melgan}. Furthermore, we adopt the feature matching loss $\mathcal{L}_{fm}$ \cite{larsen2016feature}, which is a perceptual loss for GAN training:
\begin{align}
    \mathcal{L}_{adv}\left(D\right)&=\mathbb{E}{\left[{\left(D{(y)} - 1\right)}^2+D{(G{(z_q)})}^2\right]},\\
    \mathcal{L}_{adv}\left(G\right)&=\mathbb{E}\left[\left(D{(G{(z_{q}))}-1}\right)^2\right],\\
    \mathcal{L}_{fm}\left(G\right)&=\mathbb{E}\left[\sum^{L}_{l=1}{\frac{1}{N_{l}}{||D_{l}(y)-D_{l}\left(G(z_{q})\right)||}_1}\right],
\end{align}
where $z_{q}$ denotes the quantized latent representation, $L$ is the total number of layers in discriminator $D$, $N_{l}$ represents the number of features, and $D_{l}$ extracts the feature map in the $l$-th layer of the discriminator.

\subsubsection{Auxiliary Multi-task Learning}
We introduce auxiliary tasks based on a lyrics predictor and note-pitch predictor to improve the capability of the linguistic and acoustic information in the audio codec. Each predictor takes the compressed latent representation $z_{q}$ to predict a frame-level target feature. We calculate the connectionist temporal classification (CTC) loss \cite{graves2006connectionist} between the predicted and target feature. We only apply the CTC loss to paired datasets that contain a musical score.

\subsubsection{Final Loss}
The final loss term for the audio autoencoder is defined as:
\begin{multline}
    \mathcal{L}_{gen} = \mathcal{L}_{adv}(G) + \lambda_{recon}\mathcal{L}_{recon} + \lambda_{emb}\mathcal{L}_{emb} \\
    + \lambda_{fm}\mathcal{L}_{fm}\left(G\right) + \lambda_{lyrics}\mathcal{L}_{lyrics} + \lambda_{note}\mathcal{L}_{note},
\end{multline}
where $\lambda_{*}$ is the loss weight, $\mathcal{L}_{lyrics}$ represents the CTC loss between the predicted and ground-truth lyrics, and $\mathcal{L}_{note}$ denotes the CTC loss between the predicted and ground-truth pitch IDs according to the musical instrument digital interface (MIDI) standard.

\subsection{Condition Encoder}
\label{subsec:cond_enc}
We present a condition encoder to guide the diffusion models. The condition encoder comprises a lyrics encoder, a melody encoder, an enhanced condition encoder, and a prior estimator.

\subsubsection{Lyrics Encoder}
The lyrics encoder takes a phoneme-level lyrics sequence with positional embedding as the input, and then extracts a lyrics representation. We use a grapheme-to-phoneme tool to convert the lyrics sequence into a phoneme-level lyrics sequence before feeding it into the lyrics encoder.

\subsubsection{Melody Encoder}
We introduce the melody encoder to generate a singing voice with an adequate melody from a musical score. Before using the musical score, we divide the notes into a phoneme-level note sequence. A Korean syllable generally comprises an onset, nucleus, and coda. Following the previous Korean SVS systems \cite{lee2019adversarially,choi2022melody}, we assign onset and coda to a maximum of three frames with the remainder considered as the nucleus.

Subsequently, the melody encoder extracts a melody representation from the concatenation of a note pitch, note duration, and note tempo embedding sequence with positional embedding. The note pitch sequence is transformed into the note pitch embedding. The note duration embedding sequence is represented by a fixed set of duration tokens, among which the resolution is represented by a specific note duration (e.g., the 64th note). The note tempo is calculated in beats per minute and encoded into the tempo embedding.

\subsubsection{Enhanced Condition Encoder}
The enhanced condition encoder encodes the summation of the outputs of the lyrics and melody encoders to provide a more informative condition representation $h_{cond}$. Before summing the two representations, they are expanded into the frame-level based on the note duration. In our preliminary experiments, we observed that the enhanced condition encoder effectively stabilized the pronunciation of synthesized singing voices, similar to the result in \cite{zhang2022visinger,shirahata2022period}.

\subsection{Latent Generator}
\label{subsec:latent_gen}
We adopt the latent diffusion models \cite{rombach2022high} in the latent generator to generate the latent representation of the audio autoencoder. The latent representation $\hat{z}_{0}$ is sampled using the latent diffusion models, following which the generated latent representation $\hat{z}_{0}$ is converted into the audio codec in the audio autoencoder. Furthermore, the latent representation is normalized to ease the sampling.

\subsubsection{Data-Driven Priors}
We use data-driven priors in the latent diffusion models to improve their generation abilities. Previous studies \cite{popov2021grad,lee2021priorgrad} have demonstrated that the use of data-driven priors helps approximate the trajectories between the complex data and known priors. Following \cite{popov2021grad}, we design the diffusion models to start denoising from noise close to the target $z'_{0}$, which is easier than denoising from standard Gaussian noise. We predict $\hat{\mu}$ from the condition representation $h_{cond}$ using the prior estimator of the condition encoder. We apply the negative log-likelihood loss $\mathcal{L}_{prior}$ between the normalized latent $z'_{0}$ and the predicted $\hat{\mu}$ to consider $\hat{\mu}$ as a mean-shifted Gaussian distribution $\mathcal{N}(\hat{\mu}, I)$.

\subsubsection{Latent Diffusion Models}
The diffusion process is defined using a forward stochastic differential equation (SDE) with the data-driven priors given a time horizon $t\in\left[0, 1\right]$. The forward SDE converts the normalized latent representations $z'_{0}$ into Gaussian noise:
\begin{align}
\label{eq:forward_process}
    dz'_{t} = \frac{1}{2}(\hat{\mu} - z'_{t})\beta_{t}dt + \sqrt{\beta_{t}}dW_{t},
\end{align}
where $W_{t}$ is the standard Brownian motion and $\beta_{t}$ is the non-negative pre-defined noise schedule. Its solution is expressed as:
\begin{equation}
    \begin{aligned}
        z'_{t} &= \left(I-e^{-\frac{1}{2}\int^{t}_{0}{\beta_{s}ds}}\right)\hat{\mu} + e^{-\frac{1}{2}\int^{t}_{0}{\beta_{s}}ds}z_0 \\
              &+ \int^{t}_{0}{\sqrt{\beta_{s}}e^{-\frac{1}{2}\int^{t}_{s}{\beta_{u}du}}dW_{s}},
    \end{aligned}
\end{equation}
According to the properties of It$\hat{\text{o}}$'s integral, the transition density $p_{t}{\left(z'_{t}|z'_{0}\right)}$ is the Gaussian distribution $p_{t}{\left(z'_{t}|z'_{0}\right)} \sim \mathcal{N}{\left(z'_{t};\rho_{t},\lambda_{t}\right)}$, as follows:
\begin{align}
    \rho_{t}=&\left(I-e^{-\frac{1}{2}\int^{t}_{0}{\beta_{s}ds}}\right)\hat{\mu}
    + e^{-\frac{1}{2}\int^{t}_{0}{\beta_{s}}ds}z'_0, \\
    \lambda_{t}=&I-e^{-\int^{t}_{0}{\beta_{s}ds}}.
\end{align}

We define the reverse process as an SDE solver to obtain the normalized latent representations $z'_{0} \sim p_{0}{(z')}$. We use a score estimation network $s_{\theta}$ to approximate the intractable score:
\begin{equation}
    \begin{aligned}
    dz'_{t} &= \left[\frac{1}{2}(\hat{\mu} - z'_{t}) - s_{\theta}(z'_{t}, \hat{\mu}, h_{cond}, t)\right]\beta_{t}dt\\
    &+ \sqrt{\beta_{t}}d\tilde{W}_{t}, \qquad \qquad \qquad \qquad t\in\left[0, 1\right],
    \end{aligned}
\end{equation}
where $\tilde{W}_{t}$ is the reverse Brownian motion.

Following \cite{song2020score}, we compute the expected value of the estimated gradients of the log-density of the noisy latent $z'_{t}$:
\begin{align}
\resizebox{.89\linewidth}{!}{$
\mathcal{L}_{diff}=\mathbb{E}_{z'_0, z'_t,t}\left[||s_{\theta}(z'_{t}, \hat{\mu}, h_{cond}, t) - \nabla_{z'_{t}} \log{p_{t}{(z'_{t}|z'_{0}})}||_{2}^{2}\right],
$}
\end{align}
where $ \nabla_{z'_{t}} \log{p_{t}{(z'_{t}|z'_{0}})}=-\lambda_{t}^{-1}\epsilon_{t}$ and $\epsilon_{t} \in \mathcal{N}{(0, I)}$. Furthermore, we adopt a temperature parameter $\tau$ for the data-driven prior distribution $\mathcal{N}{\left(\hat{\mu}, \tau^{-1}I\right)}$ during sampling, which helps the latent generator to maintain the quality when $\tau>1$, similar to the approach in \cite{popov2021grad}.

We jointly optimize the latent generator and condition encoder based on the following objective:
\begin{align}
    \mathcal{L}_{lg} = \mathcal{L}_{diff} + \lambda_{prior} \mathcal{L}_{prior},
\end{align}
where $\lambda_{prior}$ is the loss weight for the prior loss $\mathcal{L}_{prior}$.

\subsection{Unsupervised Singing Voice Learning Framework}
\label{subsec:unsupervised_svs}
Conventional SVS models require paired data (audio-musical score corpora) for training. Furthermore, these models cannot synthesize the singing voice of an untrained speaker without special techniques such as zero-shot adaptation. We extended our proposed model to HiddenSinger-U, an unsupervised singing voice learning framework, to mitigate the difficulty of collecting paired datasets. This framework enables the model to use unlabeled data during training. We introduce two additional encoders into the condition encoder to model the unsupervised lyrics and melody representation, as shown in Fig. \ref{fig:arch_overall} (c): an unsupervised lyrics encoder (lyrics-U encoder) and an unsupervised melody encoder (melody-U encoder). Furthermore, we employ contrastive learning in the proposed framework.

\subsubsection{Lyrics-U Encoder}
We use a self-supervised speech representation method for the linguistic information. Previous works \cite{choi2021neural,leehierspeech} have demonstrated that the speech representation from the middle layer of a self-supervised model contains phonetic information. Therefore, the phonetic information can be leveraged by extracting the self-supervised representation from the target audio. We perform information perturbation before extracting the self-supervised representation to mitigate speaker information in the target audio. The information perturbation causes the self-supervised model to focus on extracting only phonetic information. Subsequently, the lyrics-U encoder encodes the self-supervised representation into a frame-level unsupervised lyrics representation.

\subsubsection{Melody-U Encoder}
SVS models still require melody information of the target audio to synthesize singing voices. We first extract the fundamental frequency ($F0$) from the audio to extract melody information. Thereafter, we quantize the $F0$ and encode it into a pitch embedding to obscure speaker information in the target audio. Subsequently, the melody-U encoder takes the pitch embedding to extract a frame-level unsupervised melody representation.

\subsubsection{Contrastive Learning}
We observed that it is insufficient to only use the objective $\mathcal{L}_{lg}$ to optimize HiddenSinger-U owing to the gap between the paired representations (e.g., the lyrics and unsupervised lyrics representation). To maximize the agreement and penalize the dissimilarity between the paired representations, we introduce the contrastive loss \cite{chen2020simple,qian2022contentvec} for the paired data as follows:
\begin{equation}
    \begin{aligned}
        \mathcal{L}_{cont_{*}} &= \sum^{T}_{t=1}{\frac{e^{\left(\cos{(h^{(t)}_{*}, \Tilde{h}^{(t)}_{*})} / \tau_{cont} \right)}}{\sum_{\xi_{[k \neq t]}}{e^{\left( \cos{(h^{(t)}_{*}, h^{(k)}_{*})} / \tau_{cont} \right)}}}} \\
        &+ \sum^{T}_{t=1}{\frac{e^{\left( \cos{(\Tilde{h}^{(t)}_{*}, h^{(t)}_{*})} / \tau_{cont} \right)}}{\sum_{\xi_{[k \neq t]}}{e^{\left( \cos{(\Tilde{h}^{(t)}_{*}, \Tilde{h}^{(k)}_{*})} / \tau_{cont} \right)}}}},
    \end{aligned}
\end{equation}
where $\cos(\cdot, \cdot)$ calculates the cosine similarity between the pairs, $\tau_{cont}$ denotes the temperature, and $\xi_{[k \neq t]}$ represents a set of random time indices as negative samples. Following \cite{qian2022contentvec}, we randomly select several unmatched frames within each paired representation for negative samples. We apply the contrastive loss for each type of representation $h_{*}\in\left[h_{lyrics}, h_{melody}\right]$. The gap between the paired representations can be reduced by adopting the contrastive terms $\mathcal{L}_{cont_{*}}$ in the objective $\mathcal{L}_{lg}$.

\section{Experiment and Results}

\subsection{Experimental Setup}
\subsubsection{Datasets}
\label{subsec:dataset}
We trained HiddenSinger on the Guide vocal dataset\footnote{\url{https://bit.ly/3GbEUIX}} to synthesize the singing voice The Guide vocal dataset contains approximately $157.39$ hours of audio for $4,000$ paired Korean songs. We divided the audio into segments of two-bar segments to facilitate the model training, resulting in $93,127$ samples. Subsequently, we divided our dataset into three subsets: $89,186$ samples for training, $1,975$ samples for validation, and $1,966$ samples for testing.

We trained HiddenSinger-U using the Guide vocal dataset and an internal singing voice dataset containing approximately $3.30$ hours of audio for $316$ Korean songs that do not have musical scores to evaluate the unsupervised singing voice learning framework. The internal dataset was divided into three subsets: $1,130$ samples for training, $99$ samples for validation, and $97$ samples for testing. Moreover, we considered specific speakers in the Guide vocal dataset as unlabeled data during training. To train the audio autoencoder, we used the aforementioned dataset, a multi-speaker singing dataset\footnote{\url{https://bit.ly/3Q9rOkn}}, and children singing dataset \cite{choi2020children}, which contain a total of $285.1$ hours of audio for $8,781$ K-pop songs.

\subsubsection{Pre-processing}
We downsampled the audio at 24,000 Hz for training. We transformed the audio into a linear-spectrogram with 1,025 bins to train the audio autoencoder. For the reconstruction loss, we used the Mel-spectrogram with 128 bins. We grouped words into phrases and separated the phrases with the 16th rest in a text sequence for the lyrics encoder input. Subsequently, we converted the text sequence into a phoneme sequence using the grapheme-to-phoneme tool\footnote{\url{https://github.com/Kyubyong/g2p}}. We used a 64th note resolution for the note duration tokens. We used the range $[16, 256]$ for the tempo values of the tempo tokens. We extracted the self-supervised representation from the middle of XLS-R \cite{babu2021xls}, pre-trained wav2vec 2.0 \cite{baevski2020wav2vec} with 128 language dataset including Korean, as inputs for the lyrics-U encoder. Prior to the extraction, we resampled the audio at 16,000 Hz and perturbed it. We interpolated the extracted representation back to 24,000 Hz sampling rate.

\subsubsection{Training}
We trained the audio autoencoder using the AdamW optimizer \cite{loshchilov2017decoupled} with a learning rate of $2\times10^{-4}$, $\beta_{1}=0.8$, $\beta_{2}=0.99$, and a weight decay of $\lambda=0.01$. We adopted a windowed generator training \cite{donahue2020end,ren2020fastspeech,kim2021conditional} for efficiency. We randomly extracted segments of the raw waveform with a window size of 128 frames as the input for the encoder to capture the linguistic features. Furthermore, the decoder took a randomly sliced segment of the quantized latent representation $z_{q}$ with a window size of 32 frames. We used the corresponding audio segment from the ground-truth audio as the training target. Four NVIDIA RTX A6000 GPUs were used for the training. The batch size was set to 32 per GPU and the model was trained for up to 1M steps.

We jointly trained the condition encoder and latent generator using the AdamW optimizer with a learning rate of $5\times10^{-5}$, $\beta_{1}=0.8$, $\beta_{2}=0.99$, and a weight decay of $\lambda=0.01$. We randomly extracted segments of the latent representations $z_0$ with a window size of 128 frames for efficient training. We used two NVIDIA RTX A6000 GPUs for training and set the batch size to 32 per GPU. The model was trained for up to 2M steps.

\begin{figure}[t]
\centering
    \includegraphics[width=0.90\linewidth]{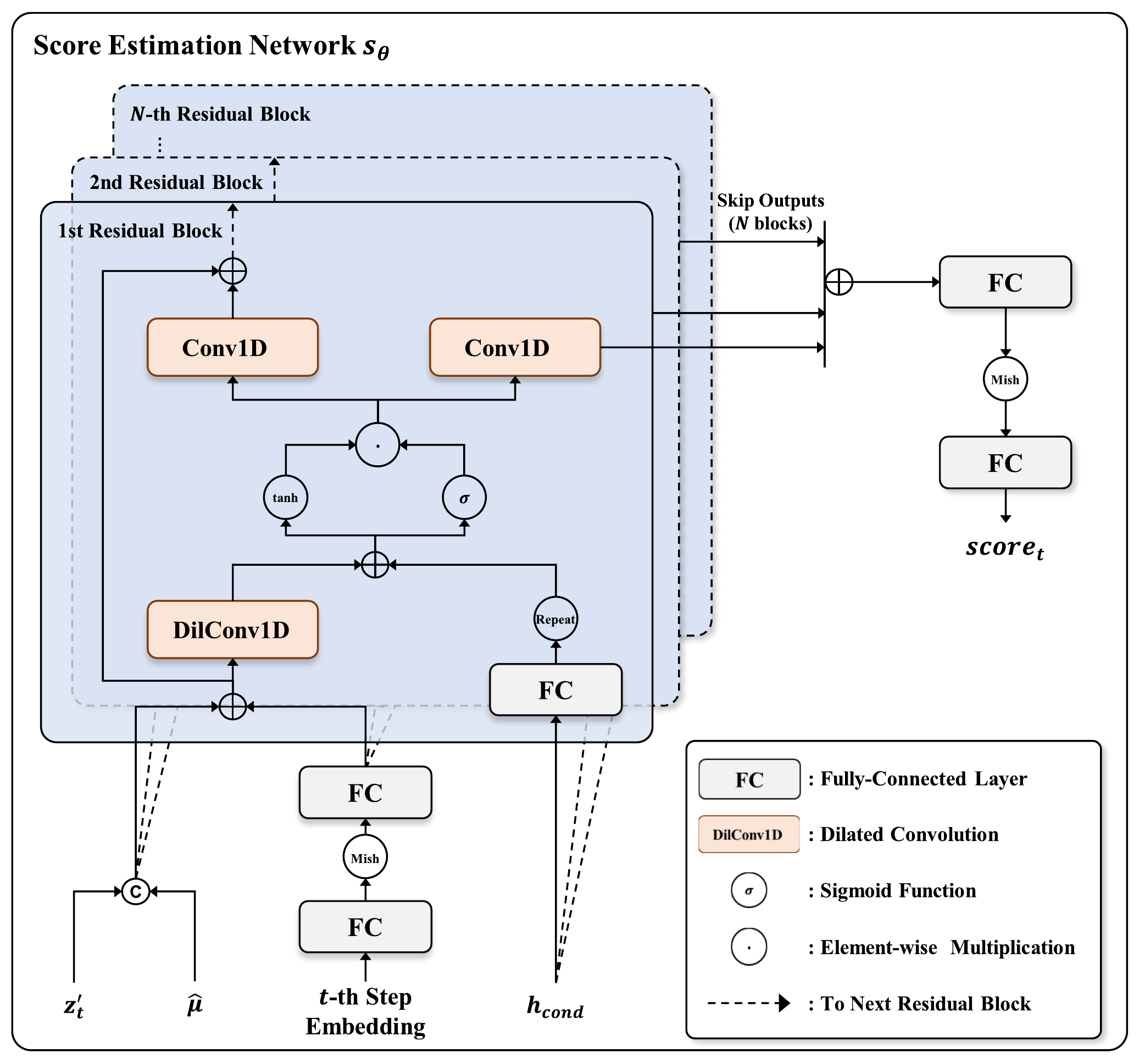}
    \caption{Architecture of the score estimation network in the latent generator}
    \label{fig:score_estimation}
    \vspace{-0.2cm}
\end{figure}

\subsection{Implementation Details}
\subsubsection{Audio Autoencoder}
The encoder comprises non-causal WaveNet residual blocks, as proposed by \cite{prenger2019waveglow}. The decoder uses a HiFi-GAN V1 generator \cite{kong2020hifi}. We implemented 30 quantizers with codebook sizes of 1,024 entries and 128 dimensions for the residual vector quantizer blocks.

\subsubsection{Condition Encoder}
The lyrics, melody, and enhanced condition encoders comprise four feed-forward Transformer (FFT) blocks \cite{ren2020fastspeech} with relative-position encoding \cite{shaw2018self} following Glow-TTS \cite{kim2020glow}. In each FFT block, we set the number of attention heads to 2, the hidden size to 192, and kernel size to 9. The prior estimator is a single linear layer.

\begin{table*}[t]
\centering
\caption{Experimental results in terms of subjective metrics and four objective metrics on the test dataset. HiddenSinger-U was trained on the same SVS dataset, of which 10\% was defined as unlabeled data.}
\resizebox{0.85\textwidth}{!}{
  \begin{tabular}{l|c|c c c c}
  \toprule
     Method & nMOS ($\uparrow$) & MAE ($\downarrow$) & Pitch ($\downarrow$)  & Periodicity ($\downarrow$) & V/UV F1 ($\uparrow$)\\
  \midrule
     GT & $4.48 \pm 0.06$ & $-$ & $-$ & $-$ & $-$\\
     HiFi-GAN (recon.) &  $4.01 \pm 0.08$ & $0.161$ & $17.893$ & $0.075$ & $0.978$\\
  \midrule
     FastSpeech 2 + HiFi-GAN & $2.27 \pm 0.08$ & $0.432$ & $55.156$ & $0.195$ & $0.939$\\
     DiffSinger + HiFi-GAN & $3.36 \pm 0.09$ & $\mathbf{0.431}$ & $45.726$ & $0.172$ & $0.949$\\
     VISinger & $3.47 \pm 0.09$ & $0.439$ & $44.441$ & $\mathbf{0.165}$ & $\mathbf{0.953}$\\
   \midrule
     HiddenSinger (Ours) & $3.80 \pm 0.08$ & $0.467$ & $\mathbf{43.247}$ & $0.172$ & $0.948$\\
     HiddenSinger-U (Ours)& $\mathbf{3.83 \pm 0.08}$ & $0.454$ & $43.536$ & $0.168$ & $0.950$\\
  \bottomrule
  \end{tabular}
}
\label{table:MOS_comparison}
\end{table*}

\subsubsection{Latent Generator}
As illustrated in Fig. \ref{fig:score_estimation}, a non-causal WaveNet-based denoiser architecture is used for the score estimation network $s_{\theta}$, similar to the architecture in \cite{kong2020diffwave,liu2022diffsinger}. We set the number of dilated convolution layers to 20, the residual channels to 256, and kernel size to 3 for the score estimation network. We set the dilation to 1 in each layer. We set $\beta_{0}=0.05$, $\beta_{1}=20$ and $T=1$ to train the latent generator and $\tau=1.5$ to sample the latent representation during inference.

\subsubsection{Unsupervised Learning Module}
The lyrics-U and melody-U encoders have the same architecture as the lyrics and melody encoders, respectively, which consist of four FFT blocks with relative-position encoding. We used the 12th layer of the pre-trained XLS-R to extract the self-supervised representation. We quantized $F0$ into 128 intervals to mitigate the speaker information.

\subsection{Subjective Metrics}
We conducted a five-scale naturalness mean opinion score (nMOS) listening test on the test dataset to evaluate the naturalness of the audio. Each audio was evaluated by 15 native Korean speakers. The subjective metrics are reported with 95\% confidence intervals in this paper.

\subsection{Objective Metrics}
We calculated the objective metrics to evaluate various types of distance between the ground-truth and synthesized audio. We considered four metrics to evaluate the SVS quality: 1) spectrogram mean absolute error (MAE); 2) pitch error; 3) periodicity error; and 4) F1 score of voiced/unvoiced classification (V/UV F1). We used the implementation of CARGAN \cite{morrison2021chunked} to evaluate the pitch, periodicity, and V/UV F1. Moreover, we provided additional objective metrics for the reconstruction quality, namely the perceptual evaluation of speech quality (PESQ) \cite{rix2001perceptual}, in Subsection \ref{subsec:autoenc}.

\subsubsection{Spectrogram mean absolute error (MAE)}
\begin{equation}
MAE=\frac{1}{T}\sum_{i=1}^{T}{|s_i - s'_i|},
\end{equation}
where $s_i$ and $s'_i$ denote the $i$-th spectrogram frame from the ground-truth and synthesized waveform, respectively. $T$ represents the frame lengths of the spectrogram.

\subsubsection{Pitch error}
\begin{align}
Pitch=\sqrt{\frac{1}{T}\sum_{i=1}^{T}{(1200 \times (\log_{2}{p_i} - \log_{2}{p'_i}))^2}},
\end{align}
where $p_i$ and $p'_i$ represent the $i$-th extracted pitch representations from the ground-truth and synthesized waveform by using torchcrepe\footnote{\url{https://github.com/maxrmorrison/torchcrepe}}, respectively. As following CARGAN, we only measure the pitch error on voiced parts in a waveform.

\subsubsection{Periodicity error}
\begin{align}
Periodicity=\sqrt{\frac{1}{T}\sum_{i=1}^{T}{({\phi}_i - {\phi}'_i)^2}},
\end{align}
where $\phi_i$ and $\phi'_i$ are the $i$-th extracted phase features from the ground-truth and synthesized waveform by using torchcrepe, respectively.

Note that the length of the synthesized and target singing voices are the same, because of the musical score that informs the duration of each note. Therefore, we do not consider time alignment, such as dynamic time warping \cite{muller2007dynamic}, to calculate objective evaluations.

\begin{figure}[t]
    \centering
    \includegraphics[width=1.0\linewidth]{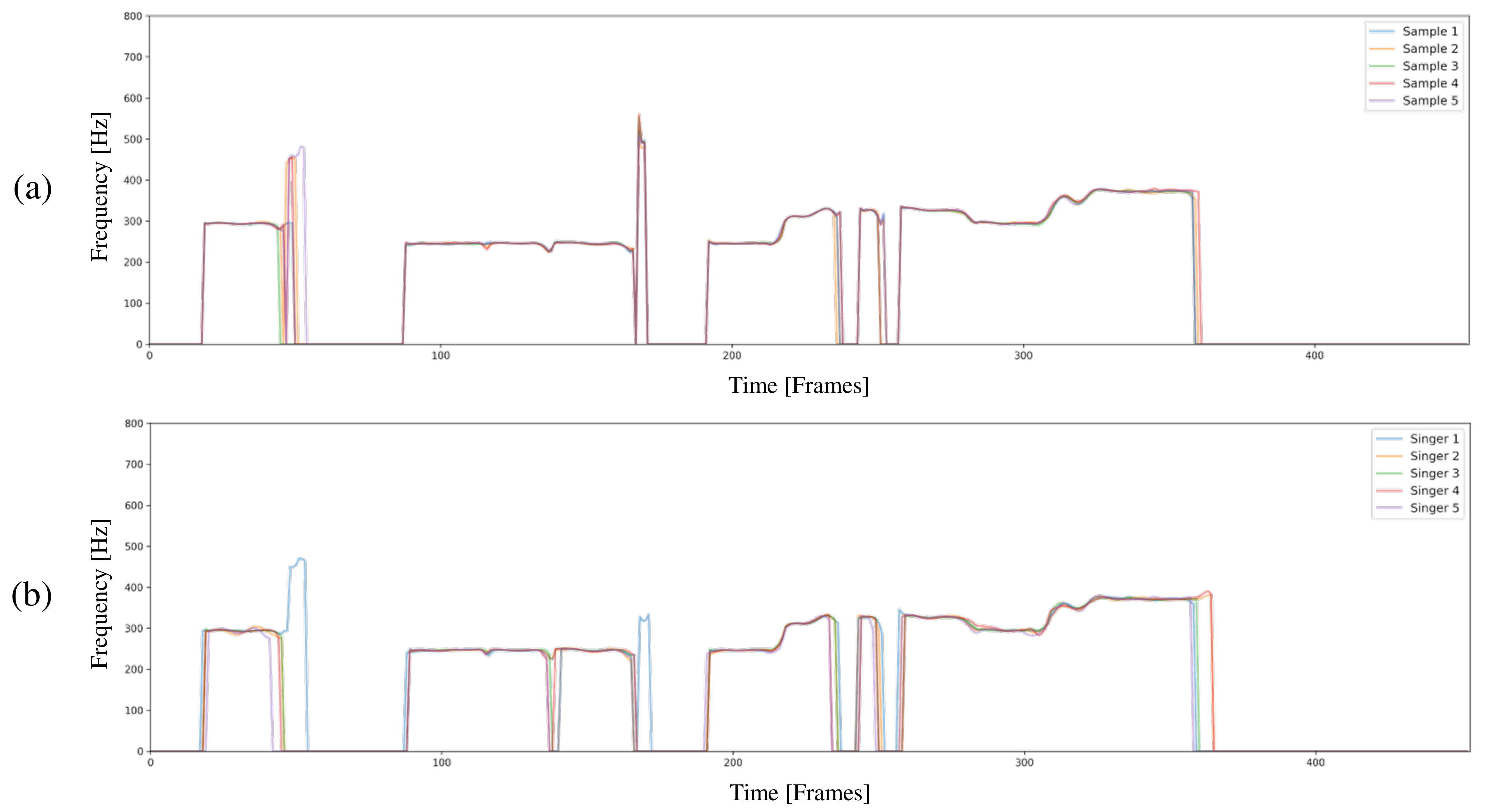}
    \caption{Visualization of generated F0 contours: (a) F0 contour variations of synthesized singing voice for five inferences with the same musical score; (b) F0 contour variations of synthesized singing voice for five speakers with the same musical score.}
    \label{fig:f0_contour}
\end{figure}

\subsection{Singing Voice Synthesis}
\label{subsec:result_svs}
We compared the audio generated by our proposed models, HiddenSinger and HiddenSinger-U, to the outputs of the following systems: 1) GT, Ground-truth audio; 2) HiFi-GAN \cite{kong2020hifi}, in which we reconstructed the audio from the ground-truth Mel-spectrogram using HiFi-GAN; 3) FastSpeech 2 \cite{ren2020fastspeech} + HiFi-GAN, in which we added a melody encoder for SVS; 4) DiffSinger \cite{liu2022diffsinger} + HiFi-GAN; and 5) VISinger \cite{zhang2022visinger}, which is an end-to-end SVS system. We trained HiddenSinger-U on the same SVS dataset, of which 10\% was defined as unlabeled data. Moreover, for fair comparisons, we trained the HiFi-GAN using the same datasets and training steps that were used to train the audio autoencoder. 

\begin{table*}[t]
\centering
\caption{Subjective and objective comparisons of reconstructed audio. Recon. indicates reconstruction. KL, reg., and RVQ denote Kullback-Leibler, regularization, and residual vector quantization, respectively.}
\resizebox{1.0\textwidth}{!}{
  \begin{tabular}{l|c|c c|c c c c}
  \toprule
     Method & MOS ($\uparrow$) & $\text{PESQ}_{WB}$ ($\uparrow$) &  $\text{PESQ}_{NB}$ ($\uparrow$) & MAE ($\downarrow$) & Pitch ($\downarrow$)  & Periodicity ($\downarrow$) & V/UV F1 ($\uparrow$)\\
  \midrule
     GT & $4.18 \pm 0.09$ & $-$ & $-$ & $-$ & $-$ & $-$ & $-$\\
  \midrule
     HiFi-GAN &  $3.67 \pm 0.11$ & $3.771$ & $3.976$ & $\mathbf{0.161}$ & $\mathbf{17.893}$ & $\mathbf{0.075}$ & $0.978$\\
     VISinger (recon.) & $3.59 \pm 0.11$ & $2.564$ & $2.986$ & $0.252$ & $25.666$ & $0.094$ & $0.974$\\
  \midrule
     Autoencoder w/o reg. & $3.86 \pm 0.09$ & $\mathbf{3.891}$ & $\mathbf{4.077}$ & $0.185$& $18.439$ & $0.076$ & $\mathbf{0.979}$\\
     Autoencoder w/ KL-reg. & $\mathbf{3.87 \pm 0.10}$ & $3.814$ & $4.045$ & $0.193$ & $19.076$ & $0.078$ & $\mathbf{0.979}$\\
     Autoencoder w/ RVQ-reg. & $3.75 \pm 0.10$ & $3.343$ & $3.675$ & $0.228$ & $18.614$ & $0.079$ & $0.978$\\
  \bottomrule
  \end{tabular}}
\label{table:audio_autoenc}
\end{table*}

As indicated in Table \ref{table:MOS_comparison}, according to the subjective audio evaluation, HiddenSinger and HiddenSinger-U outperformed the other SVS models in terms of naturalness. Moreover, our proposed models reduced the pitch error without variation predictions, such as pitch or energy prediction. These results indicate that HiddenSinger can learn accurate pitch information.

However, VISinger achieved better performance in terms of the MAE, periodicity error, and V/UV F1 score. As our proposed models generate the latent representation through stochastic iterations, the stochasticity of the models may increase the distance between the ground-truth and synthesized audio. We computed the $F0$ contour from the synthesized audio of HiddenSinger using Parselmouth\footnote{\url{https://github.com/YannickJadoul/Parselmouth}} to demonstrate the stochasticity of the models. As indicated in Fig. \ref{fig:f0_contour} (a), we performed inference five times for a speaker with the same musical score. It can be observed that HiddenSinger synthesized singing voices that contained appropriate tunes based on the musical score and variations such as intonation. As indicated in Fig. \ref{fig:f0_contour} (b), we synthesized singing voices using five different speakers and the same musical score. It can be observed that HiddenSinger generated various styles of singing voices from different speakers.

\begin{figure}[t]
    \centering
    \includegraphics[width=1.0\linewidth]{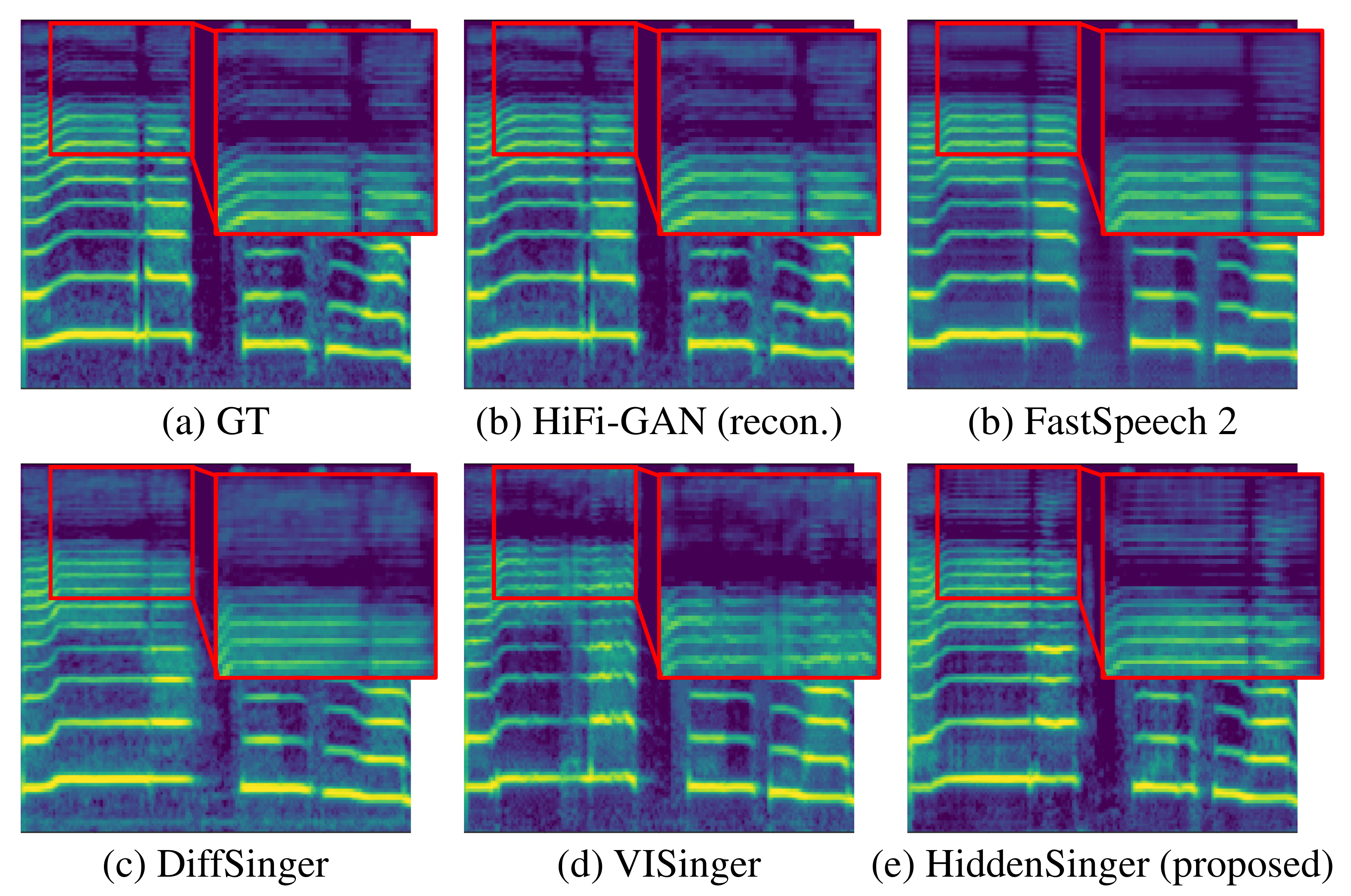}
    \caption{Visualization of generated samples with varying systems: (a) GT, (b) HiFi-GAN, (c) FastSpeech 2, (d) DiffSinger, (e) VISinger, and (f) HiddenSinger.}
    \label{fig:compare_spectrogram}
\end{figure}

Furthermore, we visualized the Mel-spectrograms of the synthesized audio to compare the models. Although the shapes of the harmonics that were synthesized by HiddenSinger differed slightly from those of the ground-truth Mel-spectrogram, the harmonics in the high-frequency band of HiddenSinger were more fine-grained than those of the other systems, as illustrated in Fig. \ref{fig:compare_spectrogram}. These results demonstrate that HiddenSinger generates high-fidelity and natural singing voices using the denoising process that can inject several variations.

\subsection{Audio Autoencoder}
\label{subsec:autoenc}
To demonstrate the performance of the audio autoencoder, we evaluated the quality of the reconstructed audio. We reconstructed the singing voice dataset used to train the VISinger for a fair comparison. As each decoder of our audio autoencoders leverages the HiFi-GAN V1 generator \cite{kong2020hifi}, they achieved similar performance to HiFi-GAN in terms of the objective evaluation metrics in Table \ref{table:audio_autoenc}. However, in terms of naturalness, our audio autoencoders achieved slightly better performance than HiFi-GAN. Moreover, the reconstruction results of the VISinger exhibited the worst performance in terms of the subjective and objective evaluation measures. These observations suggest that the end-to-end training may reduce the quality of the reconstructed audio, resulting in the upper bound of the audio generation being degraded.

\begin{table}[t]
\centering
\caption{Latent generator with different regularized audio autoencoders. $\it{LG}$ represents the latent generator.}
\resizebox{1.0\columnwidth}{!}{
  \begin{tabular}{l|c|c c c}
  \toprule
     Method & nMOS ($\uparrow$) & Pitch ($\downarrow$) & Periodicity ($\downarrow$)&  V/UV F1 ($\uparrow$)\\
  \midrule
    $\it{LG}$ w/ RVQ-reg.  & $\mathbf{4.06 \pm 0.08}$ & $\mathbf{43.247}$ & $\mathbf{0.172}$ & $\mathbf{0.948}$\\
  \midrule
    $\it{LG}$ w/o reg. & $2.97 \pm 0.08$ & $44.247$ & $0.187$ & $0.939$\\
    $\it{LG}$ w/ KL-reg. & $3.95 \pm 0.08$ & $45.825$ & $0.178$ & $0.946$\\
  \bottomrule
  \end{tabular}}
\label{table:latent_gen}
\end{table}

We evaluated the effectiveness of the different combinations of our audio autoencoder and the latent generator. We trained the latent generator separately using different regularized latent spaces. As indicated in Table \ref{table:latent_gen}, the latent generator with the RVQ-regularized autoencoder outperformed the other combinations. Furthermore, it was difficult for the latent space without regularization to generate the latent representation with the latent diffusion models. These results indicate that the RVQ-regularized latent space is more suitable for sampling targets than the KL-regularized latent space in our setting, similar to the results reported in \cite{rombach2022high}.

\begin{table*}[t]
\centering
\caption{Experimental results for unsupervised singing voice learning framework}
\resizebox{0.85\textwidth}{!}{
    \begin{tabular}{c|c c|c c c c}
        \toprule
         Unlabeled ratio & nMOS ($\uparrow$) & sMOS ($\uparrow$) & MAE ($\downarrow$) & Pitch ($\downarrow$)  & Periodicity ($\downarrow$) & V/UV F1 ($\uparrow$) \\
        \midrule 
            GT & $4.41 \pm 0.09$ & $3.79 \pm 0.05$ & $-$ & $-$ & $-$ & $-$\\
        \midrule
            $0\%$ & $\mathbf{3.81 \pm 0.11}$ & $\mathbf{3.39 \pm 0.08}$ & $0.467$ & $43.247$ & $0.172$ & $0.948$\\
            $2\%$ & $3.78 \pm 0.11$ &  $3.33 \pm 0.09$ &$\mathbf{0.454}$ & $\mathbf{42.350}$ & $\mathbf{0.165}$ & $\mathbf{0.951}$\\
            $5\%$ & $3.59 \pm 0.13$ & $2.97 \pm 0.10$ & $0.462$ & $42.819$ & $0.168$ & $0.950$\\
            $10\%$ & $3.68 \pm 0.11$ & $2.98 \pm 0.10$ & $\mathbf{0.454}$ & $43.536$ & $0.168$ & $0.950$\\
            $20\%$ & $3.60 \pm 0.12$ & $2.78 \pm 0.10$ & $0.463$ & $45.914$ & $0.172$ & $0.949$\\
            $50\%$ & $3.70 \pm 0.13$ & $2.89 \pm 0.10$ & $0.481$ & $49.084$ & $0.182$ & $0.946$\\
        \bottomrule
    \end{tabular}}
\label{table:unsupervised}
\end{table*}

\subsection{Unsupervised Singing Voice Learning Framework}
We compared the changes in the evaluation metrics according to the ratio of unlabeled data in the training dataset to verify the effectiveness of the unsupervised singing voice learning framework. We pre-defined certain speakers as unlabeled data that consisted of only audio for verification. We conducted the nMOS test to evaluate the naturalness of the audio. Moreover, we conducted a four-scale similarity MOS (sMOS) test to evaluate the voice similarity between the ground-truth and generated audio. We evaluated samples of pre-defined speakers that were considered unlabeled data in every setting, except for the 0\% and 2\% ratio settings in both MOS tests. The 0\% ratio setting represents HiddenSinger, which has been trained without the unsupervised singing voice learning framework.

It can be observed from Table \ref{table:unsupervised} that the nMOS results were statistically insignificant in most of the settings. This suggests that the unsupervised singing voice learning framework helps the model learn to synthesize a natural singing voice, regardless of changes in the unlabeled ratio. Moreover, the objective evaluations demonstrate that the proposed framework can be trained stably in every setting.

However, as shown in Table \ref{table:unsupervised}, the similarity of the synthesized singing voice decreased with an increase in the unlabeled ratio. As there were differences between the note pitch of a musical score and the $F0$ of a human speaker's singing voice, the models were trained with slightly different speaker identities due to the difference. Therefore, the human listener differentiated between the ground-truth and synthesized audio according to the difference. Although the difference degrades the similarity according to the increasing unlabeled ratio, the proposed framework is effective in synthesizing a natural singing voice with proper linguistic information and a perceptually similar speaker identity. Moreover, the contrastive terms $\mathcal{L}_{cont_{*}}$ and information perturbation aid in stabilizing the training. In particular, it is difficult to synthesize an appropriate singing voice when training is performed without the contrastive terms.

\begin{table}[t]
\centering
\caption{Ablation study of HiddenSinger. Enhanced CE represents the enhanced condition encoder in the condition encoder}
\resizebox{1.0\columnwidth}{!}{
  \begin{tabular}{l|c|c c c c}
  \toprule
     Method & nMOS ($\uparrow$) & Pitch ($\downarrow$) & Periodicity ($\downarrow$)& V/UV F1 ($\uparrow$)\\
  \midrule
     HiddenSinger & $\mathbf{3.86 \pm 0.09}$ & $\mathbf{43.247}$ & $0.172$ & $0.948$\\
  \midrule
     w/o enhanced CE & $3.24 \pm 0.10$ & $49.495$ & $0.216$ & $0.930$\\
     w/ $z_{q}$ generation & $3.73 \pm 0.09$ & $44.258$ & $\mathbf{0.170}$ & $\mathbf{0.949}$\\
     w/ standard Gaussian & $3.64 \pm 0.09$ & $45.381$ & $0.179$ & $0.947$\\
  \bottomrule
    \end{tabular}}
\label{table:abl_study}
\end{table}

\subsection{Ablation Study}
We conducted an ablation study to verify the effectiveness of each module in the proposed system. The results are presented in Table \ref{table:abl_study}. It can be observed that the subjective and objective evaluations significantly degraded with the removal of the enhanced condition encoder. Furthermore, the pronunciation of the synthesized audio was highly inaccurate without the enhanced condition encoder. Therefore, the enhanced condition encoder is necessary for the appropriate functioning of the proposed model.

We performed training on the latent generator with the audio codec $z_{q}$ as the target of the latent diffusion models. Table \ref{table:abl_study} indicates that the generation of $z_{0}$ could provide more natural audio than the generation of $z_{q}$ in the latent generator. As the RVQ blocks may refine the sampled latent representation $\hat{z}_{0}$ with residual operations, the generation of $z_0$ is superior in terms of naturalness.

Furthermore, we considered a standard Gaussian as the priors following the original denoising diffusion probabilistic models \cite{ho2020denoising}. However, the data-driven priors outperformed the standard Gaussian-based priors. This indicates that the trajectory between the data space and data-driven priors can be more stably approximate than the trajectory between the data space and standard Gaussian.

\section{Conclusions}
We have introduced HiddenSinger, a novel approach that enables the synthesis of high-quality and high-diversity singing voice audio through the integration of a neural audio codec and latent diffusion models. Our study demonstrated the efficacy of the audio autoencoder in reconstructing high-fidelity audio using low-dimensional audio codecs. Furthermore, we successfully generated latent representations conditioned on a musical score using latent diffusion models. The audio was successfully reconstructed from the generated latent representation by the audio autoencoder. We extended our model to an unsupervised singing voice learning framework that can be trained without lyrics and note information using self-supervised representation. Our latent diffusion models could be used in any speech domain, including text-to-speech and voice conversion systems. However, our model still has limitations regarding novel singing style adaptation, not voice. In future works, we will attempt to implement a zero-shot singing style transfer by adopting style-generalized generative models.

\section{Discussion}
\subsection{Broader Impact}
Recently, neural audio codecs have been used in various tasks \cite{shen2023naturalspeech,lam2023efficient}. As following concurrent works \cite{shen2023naturalspeech}, our proposed model can be extended to a text-to-speech system. Moreover, we can address the data scarcity problem by applying our unsupervised learning framework to a low-resource language.

\subsection{Social Negative Impact}
Although HiddenSinger may have practical applications such as podcasts or music generation, there is an increased risk of potential misuse of such technologies. In particular, unauthorized usage of data from web crawlers in SVS can give rise to concerns related to copyright infringement and voice spoofing. We want to emphasize that we strongly discourage the utilization of our work for any illicit or unethical purposes.

\subsection{Limitation}
Although we adopt the latent diffusion models for high-efficient latent generation, the diffusion models require a number of iterative processes to generate the representations. In the future, we will introduce the consistency models \cite{song2023consistency} to distill the teacher diffusion models for a single-step generation.  

\ifCLASSOPTIONcaptionsoff
  \newpage
\fi



\bibliographystyle{IEEEtran}
\bibliography{reference.bib}
\end{document}